\documentclass[a4paper,11pt]{article}
\usepackage{jheppub} 

\usepackage{bm}
\usepackage{amsmath}
\usepackage{slashed}
\usepackage{comment}

\newcommand{\braket}[1]{\left<#1\right>}

\title{\boldmath Chiral Symmetry Restoration 
and the Ultraquantum limit of Axionic Charge Density Waves in Weyl Semimetals}

\author[a]{Joan Bernabeu,}%
\affiliation[a]{Departamento de F\'isica de la Materia Condensada, Universidad Aut\'onoma de Madrid, Cantoblanco, E-28049 Madrid, Spain}

\emailAdd{joan.bernabeu@uam.es}

\author[a,b]{Alberto Cortijo}
\affiliation[b]{Condensed Matter Physics Center (IFIMAC), Cantoblanco, E-28049 Madrid, Spain}
\emailAdd{alberto.cortijo@uam.es}

\abstract{A new mechanism for chiral symmetry restoration at extreme high magnetic fields is proposed in the context of the Magnetic Catalysis scenario in Weyl Semimetals. Contrary to previous proposals, here we show that, at very large magnetic fields, the transverse velocity of the axion field, the phase mode of the chiral condensate $\braket{\bar{\Psi}\Psi}$, becomes effectively one-dimensional and its fluctuations destroy a possible nonzero value of this fermionic condensate. We also show that, despite of the $U(1)$ chiral symmetry not being broken at extremely large magnetic fields, the spectrum of the system is comprised by a well defined gapless bosonic excitation, connected to the axion mode, and a correlated insulating fermionic liquid that is neutral to $U(1)$ chiral transformations. When the theory is supplemented with the inclusion of dynamical electromagnetic fields, the chiral symmetry is broken again, and the conventional scenario of magnetic catalysis can be recovered.} 

\begin{document}
\maketitle
\flushbottom

\section{Introduction}

Dirac Semimetals have by now the object of study for quite sometime due to their novel features with respect to standard Fermi liquid metals. Their band structure near the so-called Dirac points is relativistic, which brings about features originally conceived in relativistic quantum field theory. These Dirac fermions can be decomponsed into fermions of left and right chirality. If their spectrum is gapless (massless in the jargon of high energy physics), the left and right-handed components are decoupled, and the Lagrangian is invariant under (global) chiral transformations at the classical level. At the quantum-level however, this symmetry is anomalously broken in the presence of parallel electric and magnetic fields.

Intimately connected with the quantum anomaly is the axion field $a(x)$. It was originally conceived as a solution to the CP problem in quantum chromodynamics (QCD). The QCD axion field couples to the anomalous contribution of the QCD gauge fields, and the minimum of the resulting effective potential for the axion field is given by the CP conserving value $a(x) = 0$.

Through the theory of Weyl semimetals, the axion field was extended to materials systems where the fermions are gapped through the presence of interactions. The phase of the order parameter now plays the role of the axion field leading to electrical conduction as the axion couples to external electromagnetic fields trough the chiral anomaly. Due to the momentum space separation of the Weyl nodes, the symmetry breaking scenario in these systems corresponds to the formation of a charge density wave (CDW) phase, and the gap in the fermion spectrum is induced by (usually strong) electronic correlations. Even if these correlations are weak, a magnetic field can catalyze the appearance of a symmetry-breaking gap. This phenomenon is known magnetic catalysis of chiral symmetry breaking, originally studied extensively in the context of QCD but also more recently in materials systems. 

The Nambu-Jona Lasinio (NJL) model was originally proposed as an effective model containing the scenario of dynamical mass generation for quarks and confinement in the context of QCD\cite{PhysRev.122.345}. One important feature displayed by the NJL model is that the system is invariant under a global continuous chiral $U(1)$ symmetry, and such symmetry is dynamically broken by the generation of quark masses. In $D=3+1$ dimensions, the phase transition to a massive phase for quarks takes place (within a mean-field level approach) at finite values $G_c$ of the coupling constant $G$ controlling the strength of the local four-fermion interaction. The model displays two parameters, the value of $G$ that has dimensions of $L^2$, and, because the NJL in four dimensions is not renormalizable, a cutoff $\Lambda$ for fermion momenta, that also allows us to interpret the NJL model as an effective theory valid at scales below $\Lambda$.

The presence of external magnetic fields changes the symmetry breaking scenario for the chiral $U(1)$ symmetry drastically\cite{gusynin1995dimensional}. In general grounds, the presence of an external magnetic fields forces a reduction of the dimensionality of the problem, from $D\to D-2$, due to the reorganization of the fermionic energy states into Landau levels. It was shown by Gusynin and coworkers that this reduction of the dimensionality carries a dynamical symmetry breaking scenario for any non-zero value of $G$. The order parameter of this symmetry breaking is the condensate $\braket{\bar{\Psi}\Psi}$ whose modulus is proportional to the value of the generated gap (mass) for fermions.

This magnetic catalysis (MC) scenario displays several remarkable features. One is related to the fact that the lowest Landau levels (LLL) of relativistic massless particles are comprised by two chiral linearly dispersing states. These two chiral states dominate the spectrum at momenta smaller than the inverse magnetic length $l^{-1}_B=\sqrt{eB}$. Also, the LLL are the only chiral states in the spectrum, so, loosely speaking, the dynamical generation of mass implies the appearance of a gap in the LLL. 
In the context of QCD, the cutoff $\Lambda_\textrm{QCD}$ is the largest scale of the problem where QCD ceases to be valid, so the assumption $\sqrt{eB}\ll\Lambda_\textrm{QCD}$ seems to be natural. However, this might not be the case in other contexts, as we discuss below.

The MC scenario has been proposed also in Condensed Matter Physics. In the context of graphene, MC implies a drastic dimensional reduction to a $0+1$ effective theory, as large enough magnetic fields renders the system in the Quantum Hall regime, where interactions become strongly relevant, as $G\sim L^{-1}$\cite{Gusynin06}. In addition, MC has also been suggested to occur in three-dimensonal Weyl semimetals\cite{roy2015magnetic}, and there is some debate from the experimental side if some Weyl semimetal candidates, like (TaSe$_4$)$_2$I, undergo a phase transition to an axionic Charge Density Wave phase in presence of an external magnetic field\cite{gooth2019axionic,Shi21}, consistent with the MC scenario, or in ZrTe$_5$\cite{Tang19}. In this particular context, the MC scenario seems to be similar to the one described in QCD, with dimensional reduction to some effective $(1+1)$ dynamics and the presence of chiral LLL. There are some important differences between Weyl semimetals and what happens in QCD, however. First, in condensed matter systems like Weyl semimetals, the relativistic description of the low energy states is an effective description \emph{within an energy regime}, and the chiral symmetry is an emergent symmetry below some given scale $\Lambda$. The presence of an ultraviolet cutoff is not only natural but necessary in condensed matter systems, which in turn forces us to interpret any property of the system from the point of view of physics in the infrared, at energy scales much smaller than this physical cutoff.

The presence of this physical cutoff poses two problems. The first one, in the context of Weyl semimetals, is that the microscopic origin of the effective coupling $G$ in the NJL model is not necessarily linked to the origin of the cutoff $\Lambda$. The typical scale of $G$ (and its bare value) is determined when establishing the microscopic origin of the local four-fermion interaction term in the model. For instance, in the context of superconductivity, this cutoff is the Debye frequency, while the NJL model arises naturally by interaction between electrons and (scalar and pseudoscalar) optical phonons\cite{PhysRevB.105.195113}. In the latter case, the coupling $G$ is related to the value of the frequency of the optical phonon mode. 

The other issue is that it might be perfectly reasonable to consider the possibility of the MC scenario in the \emph{ultra-strong quantum limit}, i.e., $\sqrt{eB}\gg \Lambda$. Although this regime is not considered in the standard approach to the NJL in QCD, in the physics of heavy ion collisions, it has been suggested that extremely large magnetic fields can be produced in such processes and the condition $\sqrt{eB}>\Lambda_\textrm{QCD}$ can be reached\cite{Shokov09}. To the best of our knowledge, this possibility was first theoretically considered in Ref.\cite{Fukushima13}. There it was argued that a new scenario could take place in the regime $\sqrt{eB}\gg\Lambda_\textrm{QCD}$, where the chiral $U(1)$ symmetry could be restored (together with a vanishing fermionic quark mass) due to the inclusion of quasi-one dimensional pion fluctuations in the effective action. The authors dubbed this scenario \emph{magnetic inhibition}. 

This scenario comprises two elements. First, the effective dynamics of the neutral pions become $(1+1)-$dimensional, as the velocity of the modes propagating perpendicularly to the magnetic field approaches zero. Second, these $(1+1)-$dimensional modes induce extra terms in the effective potential when going beyond one-loop approximation. The outcome is that, in the region of small values for the order parameter $\Delta$, this new term in the effective potential overcomes the one-loop contribution leading the solution $\Delta=0$ as the most energetically favourable. This prevents chiral symmetry breaking characterized by a fermion mass $\Delta\neq 0$, and thus inhibiting magnetic catalysis, remaining the fermionic degrees of freedom gapless, as, at the mean field level, the order parameter characterizing the chiral phase transition $\braket{\bar{\Psi}\Psi}$ is directly identified with the fermion mass $\Delta$. However, it is known from long ago\cite{kogut75,witten1978chiral} that this order parameter is proportional to the average of the Goldstone mode $\theta$ as well, $\braket{\bar{\Psi}\Psi}\propto \Delta e^{-\braket{\theta^2}/2}$, so, in the purely $(1+1)-$dimensional case,  the divergent contributions of the would-be Goldstone mode $\theta$ lead to a value for the condensate $\braket{\bar{\Psi}\Psi}=0$ although the induced fermion mass $\Delta$ might be nonzero, according to the Coleman-Mermin-Wagner-Hohenberg theorem\cite{merminwagner66,hohenberg67,coleman73}. 

In the next sections we revisit the problem of the MC when $eB\gg\Lambda^2$ as the restoration of chiral symmetry in this regime (magnetic inhibition) is not necessarily accompanied with the absence of a gap in the fermionic spectrum. We will review the MC scenario when $eB\gg\Lambda^2$ and how the dimensional reduction takes place both for fermionic and composite bosonic degrees of freedom, and bosonize the fermionic LLL to obtain an effective action description of all degrees of freedom. This bosonization scheme allows us to propose a modified scenario for the magnetic inhibition, where the chiral symmetry is restored, but (in the large $N$ limit, where $N$ is the degeneracy of LLs) there is a gapless bosonic degree of freedom, connected with the axionic degree of freedom, while the rest of the spectrum is fully gapped, corresponding to massive fermions that are neutral under chiral transformations\cite{kogut75,witten1978chiral}, in contrast to the scenario proposed in \cite{Fukushima13}. In the last sections we will discuss how the inclusion of dynamical gauge fields changes this scenario: The gauge fields also become $1+1$ dimensional in the limit $eB\gg \Lambda^2$ and hybridize with the axion field. The resulting hybrid modes render the axionic fluctuations in the order parameter short ranged, inducing again the breakdown of the $U(1)$ chiral symmetry. Despite of that, the fermionic quantum liquid remains chirality neutral and strongly correlated, again in contrast to the mean-field nature of fermions in the MC scenario. 
In the last section we comment on the possibility of observing these phenomena in Condensed Matter settings.

\section{Magnetic Catalysis and Dimensional Reduction}
The quintessential model used to study the effects of magnetic catalysis is the Nambu-Jona-Lasinio (NJL) model,
\begin{gather}\label{NJL}
    \mathcal{S}_\textrm{NJL} = \int d^4x \left\{\bar{\Psi}i\slashed{\partial}\Psi - \frac{G}{2}\left[\left(\bar{\Psi}\Psi\right)^2+\left(\bar{\Psi}i\gamma^5\Psi\right)^2\right] \right\}.
\end{gather}
In the context of Nuclear Physics, the NJL model was introduced to implement a simplistic chiral symmetry breaking mechanism and mass generation for nucleons. From the perspective of Condensed Matter Physics, the NJL can be considered to be the effective four-fermion interaction arising from the interaction of electrons and optical phonons. At energy scales smaller than the optical phonon frequency $\Omega_{ph}$, the interaction mediated by phonons become effectively local, in a very similar way as it happens in the BCS theory of superconductivity. In fact, the electron-optical phonon interaction has been suggested as a microscopic model for the formation of CDW phases in Weyl semimetals \cite{Zhang20,Kundu22,liebman23,Curtis23}. From this perspective, as mentioned in the introduction, the value of the optical phonon frequency stablishes a natural cutoff $\Lambda\sim\Omega_{ph}$ for the validity of the effective local NJL interaction. And it is conceivable that external magnetic fields can be larger than this phonon frequency, leading to the scenario studied in the present work.

In absence of external electromagnetic fields, and above a certain value for the coupling constant $G$, it is found that the fermion spectrum becomes massive \cite{PhysRev.122.345, PhysRev.124.246}. However, when an external homogeneous magnetic field $\mathbf{B} = B\mathbf{\hat{z}}$ is turned on in the system,
a chiral symmetry-breaking mass is generated for any $G>0$.
Introducing a Hubbard-Stratonovich (HS) modulus $\Delta$ and axial phase $\varphi$ fields and then integrating out the fermionic fields leads to the effective action,
\begin{equation}
    \Gamma = -i\textrm{Tr}\log\left(i\slashed{D} - \Delta e^{i\gamma^5\varphi}\right) + \int d^4x \ \frac{\Delta^2}{2G}.
\end{equation}
To find the value of the constant HS fields that minimize the energy, we find the saddle point solutions of $\frac{\delta \Gamma}{\delta \Delta}|_{\Delta, \varphi = \textrm{const}} = 0$, which leads to the gap equation,
\begin{equation}\label{gap_equation}
    0 = \frac{\Delta}{G} - \textrm{Tr}[\mathcal{D}(x,x)].
\end{equation}
Note that the $\Gamma$ is independent of any constant value of $\varphi$ owing to the global axial $U(1)$ symmetry of (\ref{NJL}), so there is no need to consider $\frac{\delta \Gamma}{\delta \varphi}|_{\Delta, \varphi = \textrm{const}} = 0$. In (\ref{gap_equation}), $\mathcal{D}(x,y) \equiv (i\slashed{D} - \Delta)^{-1}(x,y)$ is the fermion propagator in the background of a constant magnetic field in the $z$ direction \cite{PhysRev.82.664, PhysRevD.11.2124, gusynin1995dimensional}. This fermion propagator can be decomposed into a non-translation-invariant phase factor times a translation invariant component, i.e. $\mathcal{D}(x,y) = e^{i\Phi(x,y)}\tilde{\mathcal{D}}(x-y)$. The former is irrelevant for our purposes, whereas the latter can be expressed in momentum space as
\begin{equation}\label{fermion_greens_function}
   \tilde{\mathcal{D}}(p) =  i e^{-\mathbf{p}_\perp^2/(eB)}\sum_{n=0}^\infty \frac{D_n\left(p\right)}{p_\parallel^2 - \Delta^2 - 2neB},
\end{equation}
where
\begin{gather}\nonumber
    D_n(p) \equiv (-1)^n\left(\slashed{p}_\parallel + \Delta\right)\left[(1-is\gamma^1\gamma^2)L_n\left(\frac{2\mathbf{p}_\perp^2}{eB}\right)-(1+is\gamma^1\gamma^2)L_{n-1}\left(\frac{2\mathbf{p}_\perp^2}{eB}\right)\right]- \\
    - 4(-1)^n\slashed{p}_\perp L^1_{n-1}\left(2\frac{\mathbf{p}_\perp^2}{eB}\right),
\end{gather}
and the 4-momentum in the integral (\ref{greens_function_trace}) is expressed in terms of perpendicular and parallel momenta with respect to $\mathbf{B}$ as $p = (p_0,\mathbf{p}_\perp,p_3)$. The summation is done over the contribution of each Landau Level $n$ to the propagator.
Its value is, after tracing out internal degrees of freedom and evaluating at the same real-space position:

\begin{gather}\label{greens_function_trace}
    \textrm{tr}[\mathcal{D}(x,x)] = i 4\Delta \sum_{n=0}^\infty \int \frac{d^4p}{(2\pi)^4} \frac{(-1)^n\left[L_n\left(\frac{2\mathbf{p}_\perp^2}{eB}\right)- L_{n-1}\left(\frac{2\mathbf{p}_\perp^2}{eB}\right)\right]e^{-\frac{\mathbf{p}_\perp^2}{eB}}}{p_0^2 - p_3^2 - 2neB - \Delta^2}.
\end{gather}

 Integration over the perpendicular momenta provides different degeneracy factors depending on the chosen hierarchy of scales,
\begin{gather}\label{degeneracy_integral}
    \int_{|\mathbf{p}_\perp| < \Lambda} \frac{d^2p_\perp}{(2\pi)^2}(-1)^n\left[L_n\left(\frac{2\mathbf{p}_\perp^2}{eB}\right)- L_{n-1}\left(\frac{2\mathbf{p}_\perp^2}{eB}\right)\right]e^{-\frac{\mathbf{p}_\perp^2}{eB}} =
    \begin{cases}
    \frac{eB}{4\pi}\left[2 - \delta_{n0}\right] & \textrm{if } \Lambda^2 \gg eB, \\
    \frac{\Lambda^2}{4\pi}\delta_{n0}  & \textrm{if } \Lambda^2 \ll eB.
    \end{cases}
\end{gather}
When inserting this result back in (\ref{greens_function_trace}), and integrating over the energy $p_0$, one obtains different results, depending on the energy scale hierarchy. In the small field case $eB\ll \Lambda^2$, we obtain,
\begin{gather}\label{greens_function_trace_large_cutoff}
    \textrm{tr}[\mathcal{D}(x,x)] = \Delta \frac{eB}{2\pi}\int_{|p_3| < \Lambda} \frac{dp_3}{2\pi}\left[\frac{1}{\sqrt{p_3^2  + \Delta^2}}+ \sum_{n>1}\frac{2}{\sqrt{p_3^2 + 2neB + \Delta^2}}\right]= \\
    \nonumber
    = \Delta\frac{eB}{2\pi^{3/2}}\int_{\Lambda^{-2}}^\infty \frac{ds}{\sqrt{s}} \int\frac{dp_3}{2\pi}e^{-s(\Delta^2+p_3^2)}\left[1+ 2\sum_{n>1}e^{ -s(2neB)}\right] = \Delta\frac{eB}{4\pi^2}\int_{\Lambda^{-2}}^\infty \frac{ds}{s} e^{-s\Delta^2}\coth(eBs).
\end{gather}
In the last expression we have introduced the convenient parametrization in terms of Schwinger proper time. Notice that the cutoff $\Lambda$ that regularizes the $p_3$ integral in the first equality moves to the integral over the Schwinger time in the second equality. 

The result is different in the large field limit $eB \gg \Lambda^2$. In this case, the contributions to the propagator of the $n>1$ Landau levels are always suppressed in comparison with the LLL, so they can be disregarded. Integrating over the perpendicular momenta and $p_0$ similarly as before, one now obtains,
\begin{gather}\label{greens_function_trace_large_field}
    \textrm{tr}[\mathcal{D}(x,x)] = \Delta\frac{\Lambda^2}{2\pi}\int_{|p_3| < \Lambda} \frac{dp_3}{2\pi} \frac{1}{\sqrt{p_3^2  + \Delta^2}} = \Delta \frac{\Lambda^2}{4\pi^2}\int_{\Lambda^{-2}}^\infty \frac{ds}{s} e^{-s\Delta^2}.
\end{gather}
Notice that hyperbolic cotangent in (\ref{greens_function_trace_large_cutoff}) can effectively be replaced by 1 inside the proper time integral if $\Lambda^2 \ll eB$. Hence the only difference at leading between (\ref{greens_function_trace_large_cutoff}) and (\ref{greens_function_trace_large_field}) at leading order in their respective energy hierarchy is the degeneracy factor leading the integral. Finally, plugging these results into the gap equation (\ref{gap_equation}) and integrating over $\Delta$, one  recovers the expression for the effective potential dependent on the order parameter $\Delta$,
\begin{gather}\label{4D_effective_potential}
    V(\Delta) = \frac{\Delta^2}{2G} + \frac{\mathcal{N}}{4\pi}\int_{\Lambda^{-2}}^\infty \frac{ds}{s^2}e^{-s\Delta^2}\coth(eBs),
\end{gather}
where $\mathcal{N} \equiv N/S =\frac{\textrm{min}(eB, \Lambda^2)}{2\pi}$ is the LLL degeneracy $N$ per surface area $S$ of the system in the plane perpendicular to the magnetic field, and $\Lambda$ is the momentum UV cutoff for the theory (\ref{NJL}). We see then that, quite differently from the standard low-field case where the Landau Level degeneracy grows linearly with the magnetic field, for large fields this degeneracy saturates to a constant value corresponding to the highest amount of energy states that a 2D lattice with lattice spacing $a\sim\Lambda^{-1}$ can accommodate. This can be seen from the fact that in the high-field limit the magnetic length $l_{B} = (eB)^{-\frac{1}{2}} \ll a$, so the magnetic field effects for the physically-accessible LLL states are effectively coarse-grained. This point appears to be unemphasized or ignored in the literature where the $\Lambda^2 \gg eB$ limit is considered.

To recover an analytical solution for the gap equation $\delta V/\delta \Delta = 0$ in (\ref{gap_equation}), assumptions about the value of $\Delta$ must be made. In the small field limit $eB \ll \Lambda^2$, one finds that the vacuum condensate value is given by
\begin{equation}\label{expectation_value_small_field}
    \Delta_0^2 = 
    \begin{cases}
        \frac{eB}{\pi}\exp\left[- \frac{\Lambda^2}{eB}(g^{-1} - 1) - \gamma_\textrm{E}\right] & \textrm{for } g \ll 1, \\
        \Delta_{\textrm{NJL},0}^2 & \textrm{for } g > 1,
    \end{cases}
\end{equation}
where $\gamma_\textrm{E} \approx 0.577$ is the Euler-Mascheroni constant, the NJL coupling constant has been reparametrized in terms of the dimensionless constant $g \equiv \frac{G\Lambda^2}{4\pi^2}$, and $\Delta_{\textrm{NJL},0}^2$ is the mean-field solution to the NJL model for $B=0$, which can be obtained simply by setting $B=0$ in  (\ref{4D_effective_potential}). The weak coupling limit is found taking the ansatz $\Delta_0^2 \ll eB \ll \Lambda^2$, whereas the strong coupling limit is found with $eB \ll \Delta_0^2 \ll \Lambda^2$.

\begin{figure*}[ht]
    \begin{center}
    \includegraphics[width = 0.9\textwidth]{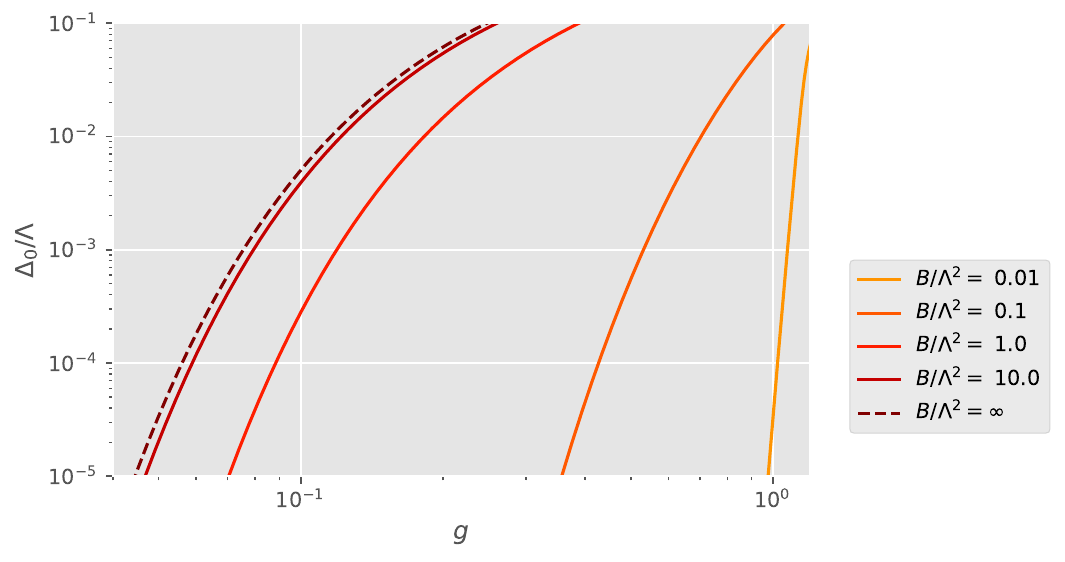}
    \end{center}
    \vspace{-9mm}
    \caption{Values of the gap $\Delta_0$ scaled with respect to $\Lambda$ as a function of the coupling $g$ for different magnetic fields. As the magnetic field becomes larger, the window for appreciable $\Delta_0$ is widened up to the asymptotic limit (\ref{expectation_value_large_field}).}
    \label{fig:gaps}
\end{figure*}

In contrast to the usual NJL model without a magnetic field \cite{PhysRev.82.664, PhysRev.122.345}, a gap is generated even for $g < 1$ \cite{gusynin1995dimensional}. Nevertheless, one can see that this gap is exponentially suppressed unless the coupling is fine-tuned to values $g \sim 1$. This can be checked in Fig.[\ref{fig:gaps}], where the curve for $\Delta_0(g)$ at $eB/\Lambda^2 = 0.01$ is seen to decrease extremly quickly for decreasing values of $g$. This is due to the $\Lambda^2/eB$ factor inside the exponential (\ref{expectation_value_small_field}), which leads to a large negative slope in a logarithmic scale presented in Fig[\ref{fig:gaps}].
 
One can now go on and consider the structure of the vacuum for ultrastrong magnetic fields. Solving the gap equation using the ansatz $\Delta_0^2 \ll \Lambda^2 \ll eB$ consistent with this limit yields
\begin{equation}\label{expectation_value_large_field}
    \Delta_0^2 = \Lambda^2\exp\left[-g^{-1} - \gamma_\textrm{E}\right].
\end{equation}
 The solution (\ref{expectation_value_large_field}) is manifestly independent of the modulus of magnetic field, whose effects are beyond the reach of the effective theory bounded to momenta and energies below the cutoff $\Lambda$. The dependence on $g$ in both the small (\ref{expectation_value_small_field}) and strong fields (\ref{expectation_value_large_field}) is the same in the relevant weak coupling regime, signalling the importance of the LLL in magnetic catalysis. Nevertheless, due to the saturation of $\mathcal{N}$ to its limiting value $\Lambda^2/(2\pi)$, the slope of the curve of $\Delta^2_0$ vs. $g$ becomes much smaller, widening the window of appreciable values for $\Delta_0$, as is seen in Fig.[\ref{fig:gaps}].

\section{Fluctuations in the phase (axion) mode.}
Similarly to other models displaying symmetry-breaking, the theory for energies below the expectation value $\Delta$ is determined by the dynamics of Nambu-Goldstone fields. Here we have only one, associated to the global $ U(1)$ chiral symmetry which we will refer hereafter to as the axion field $\theta(x)$. Its associated quanta are neutral bosons, so one may suspect that the previous reasoning of locked magnetic orbits and dimensional reduction need not apply \cite{gusynin1995dimensional}. The axion propagator is given by the expression, in terms of the fermion propagator:
\begin{equation}
    iD^{-1}_\theta(q) = -i\Delta^2\textrm{tr}[\tilde{\mathcal{D}}(k)i\gamma^5\tilde{\mathcal{D}}(k+q)i\gamma^5]. 
\end{equation}
Note how the propagator $D_{\theta}$ is translation invariant, as it is built from the translation invariant part of $\tilde{\mathcal{D}}$ in (\ref{fermion_greens_function}). This propagator can be expressed in the standard form $iD^{-1}_\theta(q) = \Delta^2Z_\theta^{-1}[q_\parallel^2 - v_\perp^2q_\perp^2]$, which itself corresponds to an effective Lagrangian (see Ref.\cite{gusynin1995dimensional}) of the form:
\begin{equation}
\mathcal{L}_\textrm{axion} \approx \Delta^2Z_\theta^{-1}\theta(x)\left(\partial^2_0- \partial^2_3- v_\perp^2\bm{\partial}^2_{\perp}\right)\theta(x).
\end{equation}
The leading factor of $\Delta^2$ is due to the choice we have made of writing the phase of the Hubbard-Stratonovich phase (the axion) as a dimensionless scalar field, $Z_\theta^{-1}$ is the axion field normalization and $v_\perp$ is the transverse velocity of the axion field (compared to the longitudinal velocity along the magnetic field).

\begin{figure*}[ht]
    \begin{center}
    \includegraphics[width = 0.9\textwidth]{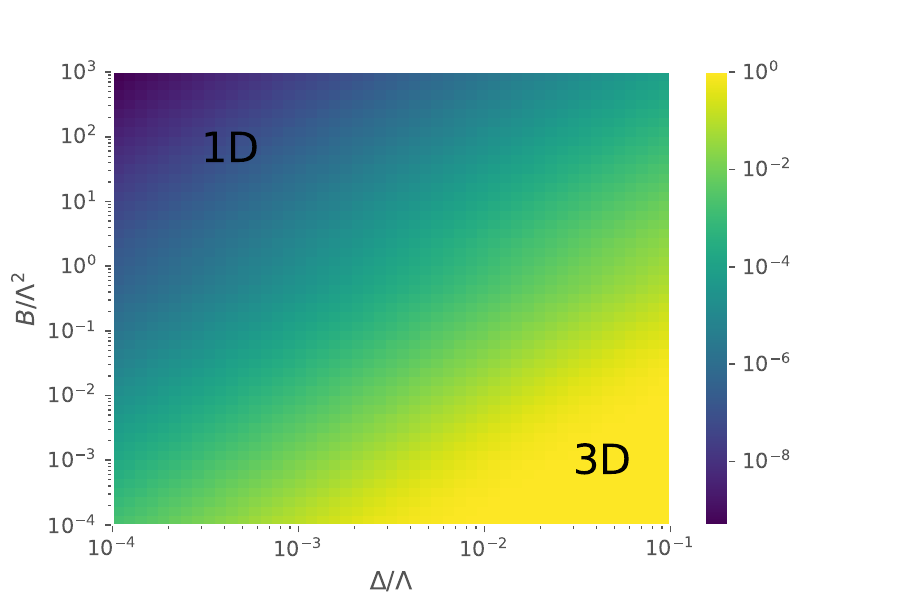}
    \end{center}
    \vspace{-9mm}
    \caption{Values of the perpendicular velocity squared $v_\perp^2$ for different gap and magnetic field values. It is seen that small gaps (i.e. weak interactions) and large fields favor the dimensional reduction from 3 to 1 spatial dimensions.}
    \label{fig:velocities}
\end{figure*}

First consider the conventional low field limit $eB \ll \Lambda^2$. For small coupling $g \ll 1$, the LLL approximation can be used. Resulting from the calculations in Appendix \ref{appendix_axion_kinetic_term}, the parameters are calculated to be, 
\begin{equation}\label{small_field_axion_parameters}
    Z_\theta^{-1} = \frac{1}{8\pi^2}\frac{eB}{\Delta^2}, \quad v_\perp^2 = \frac{\Delta^2}{eB}\log\left(\frac{\Lambda^2}{\Delta^2}\right).
\end{equation}
These results coincide with those previously cited in the literature which include all Landau levels in the weak coupling and small field limit \cite{gusynin1995dimensional,Fukushima13}. Plugging the small field, weak coupling vacuum expectation value (VEV) (Eq.(\ref{expectation_value_small_field})), one sees that the transverse velocity is proportional to a factor $\alpha\textrm{exp}[-\alpha]$ where $\alpha = \Lambda^2/(g eB)$. Due to the exponential factor, the transverse velocity is suppressed when $g\to 0$ (as it can be seen in Fig.[\ref{fig:velocities}] in the corresponding $\Delta \rightarrow 0$ limit). In fact, for vanishing coupling, the axion propagates longitudinally, as in this limit the leading prefactor for the longitudinal derivatives is $\Delta^2Z_\theta^{-1} = eB/(8\pi^2)$, which is finite. This result shows that in the absence of interactions, the theory is exactly 1+1-dimensional and can thus be bosonized exactly. Considering the strong-coupling limit $g > 1$ (see Eq.(\ref{expectation_value_small_field})), the LLL approximation is no longer valid for small fields; all Landau levels must be taken into account. At leading order in $eB/\Delta^2$, the theory amounts to that of NJL model in absence of magnetic field, which is perfectly isotropic. This result is shown in the bottom right corner of Fig.[\ref{fig:velocities}], where for large gaps and small fields $v_\perp^2 \rightarrow 1$. Hence it is seen that interactions restore the original $(3+1)-$dimensional character, as is seen in this region of Fig.[\ref{fig:velocities}], where $v_\perp^2 \rightarrow 1$ for large $\Delta/\Lambda$ (i.e. large $g$).

On the other hand, in the strong field limit $eB \gg \Lambda^2$, the parameters $Z^{-1}_\theta$ and $v^2_\perp$ take the values,
\begin{gather}\label{strong_field_axion_parameters}
    Z_\theta^{-1} = \frac{1}{4\pi^2}\frac{\Lambda^2}{\Delta^2}, \quad v_\perp^2 = \frac{2\Delta^2}{eB}\log\left(\frac{\Lambda^2}{\Delta^2}\right).
\end{gather}
 The velocity is the same, except for a factor 2, as in the small-field case in Eq.(\ref{small_field_axion_parameters}). Again, one finds that it is exponentially suppressed once the VEV (\ref{expectation_value_large_field}) is plugged in. However, now the leading factor of $\Delta^2/eB$ suppresses the velocity even for moderately large gaps. Hence the presence of a large magnetic field $eB \gg \Lambda^2$ favours the dimensional reduction, as in shown in Fig.[\ref{fig:velocities}].

Now, if the dynamics of the axion mode is effectively reduced to $(1+1)$ dimensions, then the Coleman-Mermin-Wagner theorem should inhibit any continuous symmetry-breaking, keeping the fermions massless. This is the aforementioned magnetic inhibition, which was originally studied in the small field, weak coupling regime discussed in Ref.\cite{Fukushima13}. The expectation value of the fermion condensate in the IR, $\langle \bar{\Psi}\Psi\rangle \approx - \frac{\Delta}{G} \langle e^{i\theta}\rangle$. Using that $\langle e^{i\theta}\rangle = e^{-\langle\theta^2\rangle/2}$, one can use the axion propagator
to find that for $v_\perp^2 \ll 1$,
\begin{equation}\label{condensate_scaling_with_v}
    \frac{\langle \bar{\Psi}\Psi\rangle }{\Delta/G} = \begin{cases}
        v_\perp^{\frac{1}{2}} & \textrm{for } eB \ll \Lambda^2, \\
        v_\perp^{\frac{1}{4}} & \textrm{for } eB \gg \Lambda^2.
    \end{cases}
\end{equation}
The difference in scaling exponents for the transverse velocity for each regime stems from the factor of $2$ difference in $Z_\theta$ between (\ref{strong_field_axion_parameters}) and (\ref{small_field_axion_parameters}).
However, the important observation that the order parameter $\langle \bar{\Psi}\Psi\rangle$ vanishes as it does the transverse velocity $v_\perp$ \emph{only in the limit $eB\gg\Lambda^2$}, as can be seen in Fig.[\ref{fig:velocities}].
The fact that these exponents are less than unity decreases the strength of the effects of dimensional reduction on the condensate, although they are nonetheless still in effect and the axion dynamics resembles $(1+1)-$dimensional. Then, we can go beyond the one-loop approximation performed in Ref.\cite{Fukushima13} and obtain an effective theory by bosonizing the LLL. 


\section{Lowest Landau Level Bosonization.}\label{sec:bosonization}
We follow the approach developed in Ref.\cite{PhysRevD.60.105024} and expand the fermion operator exclusively in terms of the LLL. Using a chiral basis, diagonalization of the quadratic part of Eq.(\ref{NJL}) in the background magnetic field amounts to a fermion operator with the form,
\begin{gather}
    \Psi = \begin{pmatrix}
    \Psi_L \\
    \Psi_R 
    \end{pmatrix}
    = \sum_{n=0}^{N-1} f_n(x,y)\begin{pmatrix}
        0 \\
        \psi_{L,n}(t,z) \\
        0 \\
        \psi_{R,n}(t,z)
    \end{pmatrix},
\end{gather}
where $f_n(x,y)$ is the LLL wavefunction with angular momentum quantum number $n$. Plugging this into Eq.(\ref{NJL}) and integrating over the perpendicular coordinates $(x,y)$ results in the Lagrangian $\mathcal{L} =  \mathcal{L}_0 + \mathcal{L}_\textrm{int}$ where
\begin{gather}\label{2D_full_Lagrangian}
    \mathcal{L}_0 = \sum_n \psi_{L,n}^\dagger i(\partial_0 - \partial_3)\psi_{L,n} + \psi_{R,n}^\dagger i(\partial_0 + \partial_3)\psi_{R,n}, \\
    \label{2D_interaction_Lagrangian}
    \mathcal{L}_\textrm{int} = \sum_{n,m,\bar{n},\bar{m}} g_{nm}^{\bar{n}\bar{m}} \psi_{L,n}^\dagger\psi_{R,\bar{m}}\psi_{R,m}^\dagger\psi_{L,\bar{n}},
\end{gather}
and $g_{nm}^{\bar{n}\bar{m}}$ is an interaction channel-dependent, dimensionless coupling constant, proportional to the original coupling constant $G$ in the NJL model (Eq.(\ref{NJL})). The sums are understood to run from $n=0$ to $n = N-1$. Due to the integration over the LLL wavefunctions, not all combinations of $n, m, \bar{n}, \bar{m}$ are allowed. In particular, only those that satisfy $n + m =  \bar{n} + \bar{m}$ are non-zero. Importantly, the channel satisfying the condition
\begin{equation}\label{relevant_channel}
    g_{nm}^{\bar{n}\bar{m}} = g_2\delta_ {n\bar{m}}\delta_{\bar{n}m},
\end{equation}
 survives and it turns out to be the only one needed to reproduce the effects of the magnetic catalysis transition from the perspective of bosonization. Performing a Hubbard-Stratonovich (HS) transformation on (\ref{2D_full_Lagrangian}) keeping exclusively the channel Eq.(\ref{relevant_channel}), results in a Lagrangian of the form,
\begin{equation}\label{2D_HS_Lagrangian}
    \mathcal{L}_\textrm{int} = -\Delta_2  \sum_n (e^{i\phi}\psi_{L,n}^\dagger\psi_{R,n}+ e^{-i\phi}\psi_{R,n}^\dagger \psi_{L,n}) - \frac{\Delta^2_2}{2g_2}
\end{equation}
where $\Delta_2$ and $\phi$ are the modulus and phase fields of the HS transformation. If one were to follow the mean field procedure of assuming constant values for $\Delta_2$ and $\phi$ (hence allowing $\phi$ to be rotated out), one would recover the same effective potential as (\ref{4D_effective_potential}) in the $eB \gg \Lambda^2$ limit,
\begin{gather}\label{4D_effective_potential1}
    V_2(\Delta_2) = \frac{\Delta_2^2}{2g_2} + \frac{N}{4\pi}\int_{\Lambda^{-2}}^\infty \frac{ds}{s^2}e^{-s\Delta^2_2},
\end{gather}
from which one identifies, 
\begin{equation}
    g_2 = G/S, \quad\Delta_2 = \Delta,
\end{equation}
as can be checked by plugging this identification into Eq.(\ref{2D_HS_Lagrangian}) and comparing with its 3D HS version.

Using abelian Bosonization, one can recast the theory in Eq.(\ref{2D_full_Lagrangian}) with only the channel in Eq.(\ref{relevant_channel}) into the form,
\begin{gather}\label{og_quadratic_action}
    \mathcal{L}_B = \sum_n\frac{1}{2\pi K}\left[\frac{1}{u}(\partial_0 \phi_n)^2 + u(\partial_3 \phi_n)^2\right]
    +\frac{g_2\Lambda^2}{(2\pi)^2}\sum_{n \ne m} \cos[2(\phi_{n} - \phi_{m})],
\end{gather}
where the renormalized velocity $u$ and Luttinger parameter $K$ are such that,
\begin{gather}\label{u_and_K}
    u^2 = 1 - \frac{g_2^2}{(2\pi)^2}, \quad
    K^2  = \frac{1 - \frac{g_2}{2\pi}}{1 + \frac{g_2}{2\pi}} = 1 - \frac{g_2}{\pi} + O\left(g_2^2\right),
\end{gather}
and $\Lambda$ is the UV cutoff of the theory. We can extract the axion field by defining it as
\begin{equation}\label{axion}
    \theta = \frac{1}{\sqrt{N}}\sum_{n}\phi_n.
\end{equation}
The action is then reexpressed as $\mathcal{L}_B = \mathcal{L}_\textrm{ax} + \mathcal{L}_\textrm{M}$
\begin{gather}\label{axion_lagrangian}
    \mathcal{L}_\textrm{ax} = \frac{1}{2\pi K}\left[\frac{1}{u}(\partial_0 \theta)^2 + u(\partial_3 \theta)^2\right], \\
    \label{massive_boson_lagrangian}
    \mathcal{L}_\textrm{M} = -\int d^2x\sum_{a>1}\frac{1}{2\pi K}\left[\frac{1}{u}(\partial_0 \varphi_a)^2 + u(\partial_3 \varphi_a)^2\right]
    + \frac{g_2\Lambda^2}{(2\pi)^2}\sum_{m \ne n} \cos[2\sum_{a>1}O_{mn}^a\varphi_a],
\end{gather}
where $O_{mn}^a = e_{m}^a - e_{n}^a$ and $e^a_m$ is an orthogonal matrix that satisfies $\phi_n = \sum_{a=1}^N e_n^a\varphi_a$. The definition of the axion mode, relabeled with $\theta \equiv \varphi_1$, in Eq.(\ref{axion}) implies that $e^{0}_n= N^{-\frac{1}{2}}$ for all $n$. We see that the axion field $\theta$ is a free scalar field, whereas the other fields $\varphi_a$ are scalar fields interacting through generalized sine-Gordon-like models. This is analogous to charge-spin separation in the spin$-\frac{1}{2}$ Luttinger-Liquid model \cite{PhysRevLett.33.589}, where the charge sector remains gapless if Umklapp interactions are disregarded \cite{PhysRevB.13.1272}. This is the same axion field with $(1+1)-$ dynamics discussed in the previous section, and it is completely decoupled from the other scalar fields.

We can now study the renormalization group (RG) equations for the interacting modes in Eq.(\ref{massive_boson_lagrangian}). They are similar to those of the sine-Gordon model, and no new interaction channels are generated. However, terms of the order $g_2^2$ do appear in the RG equation for $g$, in contrast to the conventional Luttinger Liquid case. Indeed, following a Wilsonian RG approach one finds that the second order correction to the action is of the form,
\begin{gather}\nonumber
\delta S_2 = -\frac{g_2^2\Lambda^4}{2(2\pi)^4}\int d^2x d^2x'  \sum_{\substack{n \ne m \\ \bar{n} \ne \bar{m}}} \langle \cos(2 O^{a}_{nm} \varphi_{a}(x))\cos(2 O^{a}_{\bar{n}\bar{m}} \varphi_{a}(x'))\rangle_> -\\
-\nonumber \langle\cos(2 O^{a}_{nm} \varphi_{a}(x))\rangle_>\langle\cos(2 O^{a}_{\bar{n}\bar{m}} \varphi_{a}(x'))\rangle_> =\\
\nonumber= -\frac{g_2^2\Lambda^4}{2(2\pi)^4}\int d^2x d^2x'  \sum_{\substack{n \ne m \\ \bar{n} \ne \bar{m}}} \cos\left[2\left(O^{a}_{nm}\varphi_{a}(x) - O^{a}_{\bar{n}\bar{m}}\varphi_{a}(x')\right) \right]\cdot\\
\label{second_order_correction_to_action}\cdot\left(e^{4\langle \varphi_a(x)\varphi_a(x')\rangle_>(\delta_{n\bar{n}} + \delta_{m\bar{m}} - \delta_{n\bar{m}} -\delta_{m\bar{n}})}-1\right)e^{-8\langle \varphi_a(0)\varphi_a(0)\rangle_>},
\end{gather}
where $\langle ... \rangle_>$ is the expected value for momentum scales in the high momentum shell to be integrated out, $[\Lambda' = \Lambda e^{-dl},\Lambda]$, where $dl$ is infinitesimal. The first equality in Eq.(\ref{second_order_correction_to_action}) is obtained after splitting the fields into low energy and high momentum components, $\varphi_a = \varphi_{a,<} + \varphi_{a,>}$. For convenience, we have left out the $<$ and $>$ labels in the second equality, but it should be understood that the fields in the cosine are in the $[0,\Lambda']$ region while the fields in the expectation value are in the renormalization shell, i.e.,
\begin{equation}
    \langle \varphi_a(x)\varphi_b(x')\rangle_> = \delta_{ab}\frac{K}{2}J_0(\Lambda |x-x'|)dl,
\end{equation}
where $J_0(u)$ is the Bessel function of the first kind. As usual , only small relative distances $|r| \equiv |x - x'|\sim \Lambda^{-1} \ll |x+x'|/2 \equiv |R|$ are relevant. In the case where strictly $n=\bar{n}$ and $m = \bar{m}$, one recovers a term proportional to the kinetic term , thus renormalizing the Luttinger parameter,
\begin{gather}
    \delta K \approx -NK^3\frac{g_2^2}{4\pi^2u^2} dl.
\end{gather}
For the case where $n=\bar{n}$ or $m=\bar{m}$, a new cosine potential with the same form as the one in Eq.(\ref{massive_boson_lagrangian}) is recovered at the lowest order in $|r|/|R|$. The coupling constant $g_2$ is therefore renormalized by
\begin{equation}
    \delta g_2 \approx  (N-2)\frac{Kg_2^2}{2\pi u} dl.
\end{equation}
At $N=2$, which corresponds to a system equivalent to 2D fermions with a $1/2$ spin degree of freedom, this term is absent as expected. Since the first order correction to $g_2$ is $O(N^0)$, then it can be ignored and the resulting equations in the large $N$ limit  are simply
\begin{gather}\label{RG_equations}
    \frac{dg_2}{dl} = NK \frac{g_2^2}{2\pi u}, \quad
    \frac{dK}{dl} = -NK^3\frac{g_2^2}{4\pi^2 u^2}.
\end{gather}
For any initial value $g_2(l_0) > 0$, the coupling becomes strong in the IR, while $K\rightarrow 0$. This is thus a variant of the Berezinskii-Kosterlitz-Thouless (BKT) transition originally derived for the XY problem, and it is a special case of a certain class of generalized multi-vertex sine-Gordon models \cite{yanagisawa2021renormalization}. The energy scale at which this transition to strong coupling occurs at lowest order in $g$ is
\begin{equation}\label{RG_scale}
    \Delta_0^2 = \textrm{min}(eB,\Lambda^2) \exp\left(-\frac{2\pi}{Ng_2}\right).
\end{equation}
Using the identifications between the parameters of the 4D and 2D theories given above, this is equivalent to the mass scales given in Eqs.(\ref{expectation_value_small_field}) and (\ref{expectation_value_large_field}). When the coupling $g_2$ becomes strong in the IR, the fields $\varphi_a$ order so that their fluctuations, represented by the $K$ parameter, are suppressed.  However, in the 2D theory the expectation value of the mass operator $\bar{\Psi}(x)\Psi(x)$ is identically zero. This is of course the consequence of assuming a strictly 1+1D theory, as taking $v_\perp = 0$ in Eq.(\ref{condensate_scaling_with_v}). This result was proven already in the context of the massless $SU(N)$ Thirring-Model in the large $N$ limit in Ref.\cite{witten1978chiral}, which corresponds to the model we are employing with the single channel (\ref{relevant_channel}), and is just the Coleman-Mermin-Wagner theorem at play. Despite this, a gap is still developed by the fermionic modes $\chi_{L/R} \equiv e^{\pm i\frac{\varphi}{\sqrt{N}}}\Psi_{L,R}$, which correspond to the interacting bosonic modes in (\ref{massive_boson_lagrangian}) and are not affected by chiral symmetry transformations. Including further interaction channels in Eq.(\ref{2D_full_Lagrangian}) would not alter this picture, as due to the $U(1)_A$-preserving form of the original NJL interaction (\ref{NJL}), there always exists a \textit{free} massless  axionic mode carrying this symmetry, as can be shown explicitly. Moreover, due to the restriction that $n+\bar{n} = m + \bar{m}$ imposed for the interaction term in Eq.(\ref{2D_full_Lagrangian}), the only other channel interaction that could possibly contribute to the bosonic kinetic term is that of $g_{nm}^{\bar{n}\bar{m}}\propto\delta_{n\bar{n}}\delta_{m\bar{m}}$. The effect of this channel is only to renormalize the velocities and Luttinger parameters of the axion, which would still remain free.

Summarizing, by using abelian bosonization, we suggest that in limit $v_\perp\to 0$, the theory described in the previous section is equivalent to a theory of a free scalar field, the axion $\theta$, and a set of $N-1$ degenerate massive bosonic fields (\ref{axion_lagrangian}-\ref{massive_boson_lagrangian}). The $N=2$ case of this model reduces to the standard spin$-\frac{1}{2}$ Luttinger Liquid model \cite{PhysRevLett.33.589} with charge-spin separation, with the massless and massive scalar field corresponding to decoupled charge and spin fields, respectively.  Eventually, the massive interacting bosonic fields can be refermionized, to obtain a degenerate set of massive, interacting fermions\cite{kogut75}. The massive character can be inferred by expanding the cosine terms in Eq.(\ref{massive_boson_lagrangian}) around minima positions. However, a complete refermionization program of the model presented in Eq.(\ref{massive_boson_lagrangian}) is beyond the scope of the present work (see, however the discussion of general $SU(N)$ models of Ref.\cite{banks1976bosonization}), as the relevant observations of being massive fermions neutral under chiral $U(1)$ transformations has been settled already. The conclusion is that in the large field limit, the axion field, as the would-be Nambu-Gostone mode associated to the chiral symmetry breaking becomes $(1+1)-$dimensional, preventing this chiral symmetry to be spontaneously broken in virtue of the Coleman-Mermin-Wagner theorem. The important observation is that, although a gapped spectrum in the fermionic sector is possible, the chiral symmetry is not broken. In this limit, exact abelian bosonization (a theory that includes fluctuations from the outset) allows us to find that the spectrum still comprises a free bosonic mode, the axion, and a set of massive fermions, although interacting and chiral $U(1)$ neutral. This is in contrast to the scenario suggested in Ref.\cite{Fukushima13}, where the inclusion of  one-loop terms in the effective potential renders the solution with massless fermions the energetically favourable scenario.

\section{Consequences from the bosonized theory: The large N limit.}
In the previous sections we uncovered how a large magnetic field $eB \gg \Lambda^2$ plays two primarily distinct roles. First, in this limit the degeneracy of the LLL is $N/S=\frac{\textrm{min}(eB, \Lambda^2)}{2\pi}$
 as stated below Eq.(\ref{4D_effective_potential}), so in the limit $eB \gg \Lambda^2$, the Landau degeneracy condition changes, not scaling with the magnetic field anymore. The other is the suppression of the transverse velocity of the Goldstone mode as can be seen in Eq.(\ref{strong_field_axion_parameters}).

This suppression of the transverse velocity has been used to justify an approach using 1D abelian bosonization of the lowest Landau level to go beyond a mean field theory (MFT) treatment when computing the effective Lagrangians in Eqs. (\ref{axion_lagrangian},\ref{massive_boson_lagrangian}) . Note that, as it can be seen in Fig. \ref{fig:velocities}, this is only valid for small gap values in the small-field case, which translates to weak couplings (in the UV). This illustrates how interactions can spoil attempts at bosonizing the Lowest Landau level. Situating ourselves in a moderately small gap (compared to the scale $\Lambda$) and strong field scenario, bosonization is justified to go beyond MFT. Having already proven how the gap scale comes about through RG methods (\ref{RG_scale}), we proceed to explore some physical consequences through the study of the CDW correlation function. 

First we review how the CDW correlation function appears in the problem at hand. Consider a generic 2-band Weyl Hamiltonian,
\begin{gather}\label{two_band_Hamiltonian}
    \mathcal{H} = \psi_{\mathbf{p}}^{\dagger} [\mathbf{\sigma}\cdot\mathbf{d}(\mathbf{p})]\psi_{\mathbf{p}},
\end{gather}
In this expression $\mathbf{d}(\mathbf{p})$ is some vector with Weyl nodes at $\mathbf{p} = \pm \mathbf{b} \equiv \pm b \hat{\mathbf{z}}$. We use the label $R$ for excitations around the right-handed node at $+b\hat{\mathbf{z}}$ and $L$ for excitations around the right-handed node at $-b\hat{\mathbf{z}}$. Assuming a homogeneous dispersion around these nodes, to lowest order in $|\mathbf{p}\pm\mathbf{b}|/(2b)$, the Hamiltonian can be split into the sum of left and right handed modes with $\psi \approx \Psi_L + \Psi_R$. The degrees of freedom are thus doubled, and, including an NJL-like interaction, the effective action in real-space shown in Eq.(\ref{NJL}) under the effects of a magnetic field in the same axis as $\mathbf{b}$ can be recovered with the replacement $i\slashed{\partial} \rightarrow i\slashed{D} + \slashed{b}\gamma^5$. In the limit $eB\ll \Lambda^2$, the the MC scenario is that left and right handed modes will couple to each other through the NJL interaction, undergoing a metallic-to insulating CDW phase transition  where the wavevector of the modulated electronic density is the distance between Weyl nodes $2b$. The axion mode corresponds to quantized fluctuations over this modulated electronic density. 

The density operator for the system is given by
\begin{gather}\label{real_space_density}
    \rho(z,\bm{x}_{\perp},t) = \rho_0(z,\bm{x}_{\perp},t) + O_\textrm{CDW}(z,\bm{x}_{\perp},t)e^{i2bz} + O_\textrm{CDW}^\dagger(z,\bm{x}_{\perp},t)e^{-i2bz}.
\end{gather}
where $\rho_0(r) \equiv \Psi^\dagger_L\Psi_L + \Psi^\dagger_R\Psi_R$  is the density of the standard 3+1D Dirac fermions, and where the CDW operator is $O_\textrm{CDW}(z,\bm{x}_{\perp},t) \equiv \Psi_R^\dagger(z,\bm{x}_{\perp},t)\Psi_L(z,\bm{x}_{\perp},t)$. After quantizing in the LLL $\Psi_\chi \equiv \sum_n \psi_{\chi,n}(z,t)f_n(\bm{x}_{\perp})(0,1)^{\textrm{T}}$, we average the density Eq.(\ref{real_space_density}) over the spatial coordinates $\bm{x}_{\perp}$ perpendicular to the magnetic field so as to define a new 1D density operator $\bar{\rho}(z,t) = \int_S 
\sum_{n,\bar{n},\chi, \bar{\chi}}\psi_{\chi,n}^\dagger \psi_{\bar{\chi},\bar{n}} f_n^*f_{\bar{n}}=\sum_{n,\chi,\bar{\chi}} \psi_{\chi,n}^\dagger \psi_{\bar{\chi},n}$ and analagously, $\bar{\rho}_0 = \sum_{n,\chi} \psi_{\chi,n}^\dagger \psi_{\chi,n}$ and $\bar{O}_\textrm{CDW} = \sum_{n} \psi_{R,n}^\dagger \psi_{L,n}$. Then we can write the effective $(1+1)-$dimensional version of Eq.(\ref{real_space_density}) in the bosonized language as,
\begin{gather}\nonumber
    \bar{\rho}(z,t) = \sum_n -\frac{1}{\pi}\partial_z\phi_n + \frac{\Lambda}{2\pi}e^{2ibz - 2i\phi_n} + \textrm{h.c.}=\\
    = -\frac{\sqrt{N}}{\pi}\partial_z\theta   + \sum_n \frac{\Lambda}{2\pi}e^{i 2bz - i2\sum_{a>1}e_n^a\varphi_a -\frac{i2}{\sqrt{N}}\varphi} + \textrm{h.c.}.
\end{gather}
The derivative term, which corresponds to $\bar{\rho}_0 \equiv -\frac{\sqrt{N}}{\pi}\partial_z\theta$, is the contribution from the axion through the chiral anomaly \cite{roy2015magnetic}, whereas the second term corresponds to the contribution from the remaining modes, $\bar{O}_\textrm{CDW}$. Each of them leads to different contributions to the density-density correlation function, 
\begin{equation}
\langle \bar{\rho}(z,t)\bar{\rho}(0)\rangle = \langle\bar{\rho}_0(z,t)\bar{\rho}_0(0)\rangle + \langle \bar{O}^\dagger_\textrm{CDW}(z,t)\bar{O}_\textrm{CDW}(0)\rangle e^{-i2bz} + \textrm{ h.c.}.\label{density_correlation}
\end{equation}
As already mentioned, the first term in Eq.(\ref{density_correlation}) is the contribution from the axion mode, and the rest comes from the remaining modes. Due to the fact that the axion field is decoupled from the massive bosonic fields, it can be factorized,
\begin{eqnarray}\label{CDW_factorization}
   \nonumber \langle \bar{O}^\dagger_\textrm{CDW}(z,t)\bar{O}_\textrm{CDW}(0)\rangle = \langle e^{\frac{2i}{\sqrt{N}}\theta(z,t)}e^{-\frac{2i}{\sqrt{N}}\theta(0)}\rangle\cdot\\ \cdot\frac{\Lambda^2}{4\pi^2}\sum_{n,m}\langle e^{2i\sum_{a>1}e_n^a\varphi_a(z,t) }e^{- i2\sum_{a>1}e_m^a\varphi_a(0)}\rangle.
\end{eqnarray}
The sum over $n,m$ is a constant (coordinate-independent) in the low-energy limit, as there the RG equations in Eq.(\ref{RG_equations}) dictate that the bosonic fields different from the axionic mode order at the bottom of the potential in Eq.(\ref{massive_boson_lagrangian}) and become massive. To estimate their contribution to Eq.(\ref{CDW_factorization}), we use the argument given in the last paragraph of the previous section. At low energies, the massive sector can be refermionized into self-interacting fermion fields with mass $M=\Delta_0$ given in Eq.(\ref{RG_scale}). This can be seen by expanding the cosine to quadratic order in the bosonic fields $\varphi_a$, which are seen to decouple. This is the same limit that would be obtained in the IR after bosonizing $N-1$ self-interacting fermions $\chi_n$. Applying a MFT approach to these modes in the $eB\gg \Lambda^2$ limit, which is now allowed since they are decoupled to the massless bosonic mode, these new fermionic modes would obtain a mass $\Delta_0$. Therefore, to leading order in $N$, the summations in Eq.(\ref{CDW_factorization}) lead to:
\begin{gather}\label{massive_fermion_correlation}
    \nonumber \sum_{n,m}\langle\bar{\chi}_n(z,t)(1-\gamma^5)\chi_n(z,t)\bar{\chi}_m(0)(1+\gamma^5)\chi_m(0)\rangle\to \\ \to\frac{N^2\Delta_0^2}{(2\pi)^2}\left[\log\left(\frac{\Lambda^2}{\Delta_0^2}\right)\right]^2 \approx \frac{\Delta_0^2}{g^2}.
\end{gather}
The right-hand side is the expected contribution from the MFT. The contribution from the axion field is easily found by extrapolating the well-known $N=1$ correlation function (see e.g. Ref.\cite{giamarchi2003quantum}) to large $N$, and at finite temperature $T$ the correlation function reads,
\begin{gather}\label{CDW_correlation_function}
    \langle \bar{O}_\textrm{CDW}^\dagger(z,t)  \bar{O}_\textrm{CDW}(0)\rangle = \frac{\frac{\Delta_0^2}{g^2}\left(\frac{\pi T}{u\Lambda}\right)^{\frac{2K}{N}}}{\left[\textrm{sinh}\left(\pi T\left(\frac{z}{u}-t +i\epsilon s\right)\right)\textrm{sinh}\left(\pi T\left(\frac{z}{u}-t +i\epsilon s\right)\right)\right]^{\frac{K}{N}}},
\end{gather}
where $s \equiv \textrm{sign}(t)$ and $\epsilon$ is an infinitesimal number. One can thus calculate the contribution to the retarded charge susceptibility, i.e.  $\langle \bar{O}^\dagger_\textrm{CDW}(z,t)\bar{O}_\textrm{CDW}(0)\rangle^R e^{-i2bz} + \textrm{ h.c.}$, in frequency ($q_0$) and momentum ($q_3$) space,
\begin{gather}
{\textstyle
    \chi^{R}_{\textrm{CDW}}(q_0,q_3) \propto
    \frac{1}{q_0^2 - u^2|q_3-2b|^2 - \left(\frac{2\pi K T}{N }\right)^2 + i\frac{4\pi K T}{N }q_0} + \frac{1}{q_0a^2 - u^2|q_3+2b|^2 - \left(\frac{2\pi K T}{N }\right)^2 + i\frac{4\pi K T}{N}q_0}},
\end{gather}
which has four poles at
\begin{equation}\label{poles}
    \omega_{\pm,\pm} = - i\frac{4\pi K T}{N} \pm u|q_3 \pm 2b|.
\end{equation}
In the homogeneous limit where $q_3 \sim 0$, the limit $N\rightarrow\infty$ effectively destroys the thermal broadening of the linewidth and one recovers the zero temperature result. The charge susceptibility $ \chi^{R}_{\textrm{CDW}}$ can be inferred from inelastic light scattering experiments \cite{Schulz90,Deveraux07,Orignac12}.

Besides fluctuations in the density, one could also choose to study the dynamics of fluctuations present in the current. However, the current with respect to a spatially homogeneous electric field, as is prevalent in experimental setups, can be studied just for the $q\sim 0$ contribution,
\begin{equation}\label{current_operator}
    j(z,t) = \sum_{n}\frac{1}{\pi}\partial_0\phi_n = \frac{\sqrt{N}}{\pi}\partial_0\theta.
\end{equation}
This quantity is independent of the effects of the CDW fluctuations, as it corresponds to same current predicted by the 3D chiral anomaly \cite{roy2015magnetic}, which, when translated to the 1D theory, will lead to the 1D anomaly $\mathcal{L}_\textrm{an} \propto \sqrt{N} E \theta$ (see e.g. \cite{PhysRevLett.126.185303}), where $E$ is the component of the electric field parallel to the magnetic field. Despite of that, it is interesting how the dynamics of the current-current correlation function, that is, the conductivity, is affected by the change in the degeneracy $N$ from the limit $eB\ll \Lambda^2$ to $eB\gg \Lambda^2$. 

The calculation of the conductivity for the 1D system is equivalent to calculating the conductivity along the axis of the magnetic field in the 3D system, i.e. the magneto-conductivity $\sigma_m\propto \langle j(z,t) j(0)\rangle$. Equation (\ref{current_operator}) implies then that  $\sigma_m \propto N$. In the $eB\ll \Lambda^2$ limit, this reproduces the well known result $\sigma_m \propto eB$ \cite{son2013chiral}. However, since in the large field $eB\gg \Lambda^2$ limit considered here the degeneracy $N$ is no longer dependent on the magnetic field strength, one can expect that, for sufficiently large magnetic fields, the magneto-conductivity as a function of the magnetic field should saturate to a constant value,as it was already discussed that this is due to the saturation of available states provided by the system in the plane perpendicular to the magnetic field. The transition from the MC scenario at $eB\ll \Lambda^2$ to the scenario at $eB\gg \Lambda^2$ where chiral symmetry is restored can be experimentally traced by measuring such magnetic field saturation of the magnetoconductivity.

\section{1+1D Axionic Polaritons: Restoration of Chiral Symmetry Breaking.}
Conservation of chiral symmetry in the ultraquantum limit fundamentally depends on the axion being massless as a mass would give $\langle \bar{\Psi} \Psi \rangle \propto \langle e^{i\theta}\rangle =  e^{-\langle\theta^2\rangle/2}$ a finite, non-zero value. It is a well known fact however that the axion can hybridize with the electric field modes of the media, through the chiral anomaly, leading to axionic polaritons \cite{Li09Dynamical}, where one of the hybridized modes is gapped, but the other resulting mode remains gapless. One might thus expect that the IR fluctuations of the gapless mode would uphold the Coleman-Mermin-Wagner theorem. As we shall shortly see, this is not the case.

The standard derivation of this result only takes into account the mixed vector-axial current response of fermionic Landau Levels, i.e. the chiral anomaly, but neglects the vector-vector response, i.e. the photon polarization. Indeed, calculating the polarization response of the system in the LLL and IR limit \cite{SHABAD1975166, miransky2015quantum}  yields $\Pi^{\mu\nu} \propto \frac{\textrm{min}(\Lambda^2, eB)}{\Delta^2}(q_\parallel^{\mu}q_\parallel^{\nu} - q^2_\parallel\eta^{\mu\nu}_\parallel)$ (for the polarization including all Landau levels, see \cite{Batalin71Photon,Tsai74Vacuum}). This in turn implies that the electromagnetic field velocity for modes propagating perpendicular to $\mathbf{B}$ will be suppressed. Therefore, we can describe the system with an effective 1+1D Lagrangian that mixes the 2-vector potential $A^\mu = (A^0, A^3)$ and the axion field $\theta$. In the $A^0 = 0$ gauge, it reads
\begin{gather}\label{lagrangianPolariton}
    \mathcal{L} = -\frac{1}{2}A^{3}(\Box_A + c^2\partial_3^2) A^{3}  -\frac{1}{2}\theta\Box_\theta\theta + \kappa \theta \partial_0A^3,
\end{gather}
where $\Box_A \equiv \partial_0^2 - c^2\partial_3^2$ and $\Box_\theta \equiv \partial_0^2 - v^2\partial_3^2$, and $\kappa$ is a constant that absorbs the anomaly coefficient and the field renormalizations. In frequency ($q_0$) and momentum ($q_3$) space, it can be expressed as
\begin{gather}\label{lagrangianPolaritonMomentum}
    \mathcal{L} = \frac{1}{2} \begin{pmatrix} A^3_{-q} & \theta_{-q} \end{pmatrix} \begin{pmatrix}
        q_0^2 & i\frac{\kappa}{2}  q_0 \\ -i\frac{\kappa}{2} q_0 & q_0^2 - v^2q_3^2
    \end{pmatrix} \begin{pmatrix} A^3_{q} \\ \theta_{q} \end{pmatrix} =  \frac{1}{2} \begin{pmatrix} \phi_{0,-q} & \phi_{1,-q} \end{pmatrix} \begin{pmatrix}
       \lambda_0(q) & 0 \\ 0 & \lambda_1(q)
    \end{pmatrix} \begin{pmatrix} \phi_{0,q} \\ \phi_{1, q} \end{pmatrix}
\end{gather}
where $\lambda_0(q) = \frac{1}{2}\left(2q_0^2 - v^2q_3^2 + \sqrt{v^4q_3^4 + \kappa^2q_0^2}\right)$ and $\lambda_1(q) = \frac{1}{2}\left(2q_0^2 - v^2q_3^2 - \sqrt{v^4q_3^4 + \kappa^2q_0^2}\right)$. These eigenvalues correspond to the $\omega_0 = 0$ and $\omega_1 = \sqrt{v^2q_3^2 + \frac{\kappa^2}{4}}$ modes and the new fields $\phi_0$ and $\phi_1$, respectively. The original fields $(A^3,\theta)$ can be now written in terms of these hybrid modes $(\phi_0,\phi_1)$ as
\begin{gather}
\begin{pmatrix} A^3_{q} \\ \theta_{q} \end{pmatrix} =
\begin{pmatrix}
    \cos \alpha_q & -i\sin\alpha_q \\
    -i\sin\alpha_q & \cos \alpha_q
\end{pmatrix}
\begin{pmatrix} \phi_{0,q} \\ \phi_{1, q} \end{pmatrix},
\end{gather}
where
\begin{gather}
    \cos\alpha_q = \frac{1}{\sqrt{2}}\left[1 + \frac{v^2q_3^2}{\sqrt{v^4q_3^4 + \kappa^2q_0^2}}\right]^{\frac{1}{2}},
    \sin\alpha_q = \frac{1}{\sqrt{2}}\left[1 - \frac{v^2q_3^2}{\sqrt{v^4q_3^4 + \kappa^2q_0^2}}\right]^{\frac{1}{2}}.
\end{gather}
This implies that the expectation value for the axion field $\theta$ reads,
\begin{gather}\label{expectation_value_polariton}
    \langle \theta^2\rangle = \int \frac{d^2q}{(2\pi)^2} \frac{i\sin^2\alpha_q}{\lambda_0(q)} + \int \frac{d^2q}{(2\pi)^2} \frac{i\cos^2\alpha_q}{\lambda_1(q)} = \int \frac{d^2q}{(2\pi)^2} \frac{i}{q_0^2 - v^2q_3^2 - \kappa^2/4}.
\end{gather}
The contribution to $\langle \theta^2\rangle$ of the two hybrid modes combine to produce the structure of the pole for a single gapped boson. Since the expectation value only depends on the gapped pole, the integral is clearly IR-finite and therefore, $\langle \bar{\Psi} \Psi \rangle \propto  e^{-\langle\theta^2\rangle/2}$ is non-zero. We conclude thus that chiral symmetry is therefore broken when taking into account the internal electromagnetic degrees of freedom of the system in the approximately 1+1D ultraquantum limit. The fact that (\ref{expectation_value_polariton}) only depends on the gapped pole is a result exclusive to this 1+1D approximation. Indeed, an analogous calculation for the 3+1D case \cite{FERRER2023116307} would show a dependence on the pole of the gapless mode. Of course, in the 3+1D case IR fluctuations are not an impediment to chiral symmetry-breaking to begin with.

In a similar fashion to the condensate, in the 1+1D case considered here, the conductivity only depends on the massive pole. Indeed, when an external electric field $E(x)$ parallel to the magnetic field is applied to the system, one can solve the EOM for the two hybrid modes in the Lagrangian (\ref{lagrangianPolaritonMomentum}). As it happened when discussing Eq.(\ref{expectation_value_polariton}), both modes contribute to the current coming from the axion and conspire to generate the same gapped pole structure in the conductivity $\sigma(q)$:
\begin{gather}
    j_q = iq_0 \theta_q = iq_0\left[\frac{\sin^2\alpha_q}{\lambda_0(q)} + \frac{\cos^2\alpha_q}{\lambda_1(q)}\right]\kappa E_q =  \frac{iq_0\kappa}{q_0^2 - v^2q_3^2 - \kappa^2/4}E_q\equiv\sigma_m(q)E_q.
\end{gather}
For spatially homogeneous fields, $E_q\sim E(q_0)\cdot\delta(q_3)$, the magneto-conductivity $\sigma_m(q_0)$ goes to zero in the DC limit, $q_0 = 0$, as expected from a fully insulating system. However, important aspect to note is that the axionic polariton gap $\kappa^2/4$ in the ultraquantum limit $eB \gg \Lambda^2$ will be independent of the value of the magnetic field, as all the quantities involved in this limit, i.e. the anomaly coefficient and the field renormalization parameters, will also be independent of $B$.

It is interesting to comment here that both mechanisms of chiral symmetry restoration discussed in the present work, and the one proposed in Ref.\cite{Fukushima13} assume that the axion, as the would-be Goldstone mode associated to the chiral symmetry breaking, is massless, irrespective to the value of the transverse velocity $v_{\perp}$. It is known that Quantum Electrodynamics in $(1+1)-$dimensions (Schwinger model) describing relativistic $(1+1)-$ fermions coupled to dynamical $U(1)$  $(1+1)-$dimensional gauge fields exhibits chiral symmetry breaking despite of the low dimensionality \cite{Schwinger62}. It has been recently suggested that the Schwinger model, when considered together with the NJL model, induces a gap in the axion mode, initially identified with a Chiral Magnetic Wave \cite{Kharzeev11,Mottola21,Mottola23} (see however, Ref.\cite{Shovkovy19}). This mechanism strongly resembles to the one proposed in the past in the context of CDW to obtain massive phasons \cite{Lee78} and it has been invoked to explain recent experimental results of photo-emission  spectroscopy on the axionic CDW candidate material (TaSe$_4$)$_2$I, that points along the presence of a gapped phason mode \cite{Kim23}. While the gap in the axion due to the hybridization with electromagnetic modes has no direct impact in the mechanism of Magnetic Catalysis (the chiral symmetry breaking takes place irrespective the dynamics of the Goldstone boson, as fluctuations of the Goldstone mode do not destabilize the nonzero value of the condensate) it is all important in our scenario, as the net contribution of axion-electromagnetic hybrid modes renders the fluctuation of the order parameter finite, going from a restoration of the chiral symmetry breaking to its breakdown.

\section{Conclusions}

As the role of the ultraviolet cutoff $\Lambda$ is essential in the scenario proposed in previous section, we need to provide an estimation of its value. A customary choice in Condensed Matter Physics is just to assume $\Lambda$ to be proportional to the inverse lattice spacing, i.e. $\Lambda = \kappa a^{-1}$, where $\kappa$ is some factor of proportionality, often taken as $2\pi$ (but the reader should recall that $\Lambda$ is the value of the momenta where the linear approximation in the band structure ceases to be valid, it might well be any smaller value). For a benchmark value of $a = $ 5 \AA, the condition $\Lambda^2 \sim eB$ (this is the lower bound for magnetic fields where to expect the theory presented here to apply) implies a magnetic field of $B \sim 2500\kappa^2$ T, hence a prefactor of $\kappa\sim 0.1$ would suffice in achieving an experimentally attainable $B = 25$T. 
It is worth noting that the quantum limit can be reached in current condensed matter systems like ZrTe$_5$ at magnetic fields below 3 T, driving the system to several phase transitions\cite{Tang19} (in the case of HfTe$_5$, a metal-insulator transition has been reported at crital magnetic fields of the order of $7$ T \cite{Wang20}). It is hence conceivable that for larger magnetic fields, the conditions for the scenario presented here could be reached in similar systems.

In summary, we have developed a theory for the $U(1)$ chiral symmetry restoration and CDW phases in Weyl semimetals in the limit $eB\gg \Lambda^2$. Despite the symmetry being restored, the system undergoes a phase transition to a correlated insulating phase with modulated charge density, and where the Goldstone bosons do not disperse in the directions transverse to the magnetic field, allowing us to employ bosonization techiques to study the effective theory including fluctuations of the Goldstone modes from the outset.  This CDW is characterized by an axionic mode that couples to external electric fields through the one-dimensional version of the chiral anomaly, and a set of degenerate massive fermions that are neutral to the $U(1)$ chiral symmetry, in contrast to the original electronic degrees of freedom. Finally, we show that when the dynamic electromagnetic degrees of freedom of the system are taken into account, the symmetry restoration is ultimately impaired.

\section{Acknowledgements}
J.B. is supported by FPU grant FPU20/01087. A.C. acknowledges financial support from the Ministerio de Ciencia e Innovaci\'on through the grant PID2021-127240NB-I00 and the Ram\'on y Cajal program through the grant No. RYC2018- 023938-I. J. B. and A. C. acknowledge discussions with K. Landsteiner, F. Peña Benitez, M. N. Chernodub, M. A. H. Vozmediano, Mireia Tolosa-Sime\'on, Bilal Hawashin, and Michael M. Scherer. 

\appendix
\section{Appendix: Axion Kinetic Term}\label{appendix_axion_kinetic_term}
We are interested in calculating the matrix,
\begin{gather}\label{Fmunu}
    F^{\mu\nu} = -\frac{i}{2}\int\frac{d^4k}{(2\pi)^4}\textrm{tr}\left[\tilde{D}i\gamma^5\frac{\partial^2\tilde{D}}{\partial k_\mu\partial k_\nu }i\gamma^5 \right],
\end{gather}
which corresponds to the kinetic coefficient of the axion Lagrangian,  $\mathcal{L}_\textrm{axion} = \frac{\Delta^2}{2}F^{\mu\nu}\partial_\mu\theta\partial_\nu\theta$ \cite{gusynin1995dimensional}.
This quantity is only nonvanishing if $\mu = \nu$. We will perform this calculation in the LLL approximation, i.e. taking into account only the $n=0$ term inside the summation in (\ref{fermion_greens_function}). Strictly speaking this approximation is only valid in the $\Lambda^2 \ll eB$ limit, but it also reproduces well the results in the weak coupling limit of $eB \ll \Lambda^2$ \cite{gusynin1995dimensional, Fukushima13}. We consider first the value for the coordinates perpendicular to the magnetic field, i.e. $\mu=\nu = 1$ or $\mu = \nu = 2$. Using that $F_\perp \equiv -F^{11} = -F^{22}$ and one finds that
\begin{gather}\nonumber
    F_\perp =  -\frac{F^{11} + F^{22}}{2} =  8i\left\{\int\frac{d^2k_\parallel}{(2\pi)^2}\frac{1}{k_\parallel^2 + \Delta^2}\right\}\left\{\int \frac{d^2k_\perp}{(2\pi)^2}\frac{eB-k_\perp^2}{(eB)^2}e^{-2\frac{k_\perp^2}{eB}}\right\}=
    \\ = \frac{2}{\pi^2}\left\{ \int_0^\Lambda \frac{k_\parallel dk_\parallel}{k_\parallel^2 + \Delta^2}\right\}\left\{ \int_0^\Lambda \frac{k_\perp dk_\perp}{eB}\left(1-\frac{k_\perp^2}{eB}\right)e^{-2\frac{k_\perp^2}{eB}}\right\}
\end{gather}
where in the second line we have integrated over the angular variables and Wick-rotated the integral over parallel momenta $k_\parallel$. Performing the integral in the low and high field limits results in $F^{\mu\nu}_\perp = \eta^{\mu\nu} F_\perp$, where
         \begin{equation}\label{Fperpendicular}
    F_\perp =
    \begin{cases}
        \frac{1}{8\pi^2}\log\left(\frac{\Lambda^2}{\Delta^2}\right) & \textrm{if } eB \ll \Lambda^2 \\
        \frac{1}{2\pi^2}\frac{\Lambda^2}{eB}\log\left(\frac{\Lambda^2}{\Delta^2}\right) & \textrm{if } \Lambda^2 \ll eB.
    \end{cases}
\end{equation}
For the term corresponding to the parallel coordinates, i.e. $\mu = \nu = 0$ or 3, one can define as before $F_\parallel = F^{00} = - F^{33}$. Proceeding analogously as for the perpendicular component, one finds
\begin{gather}\nonumber
    F_\parallel = \frac{F^{00}-F^{33}}{2} =\frac{2}{\pi^2}\left\{\int_0^\Lambda \frac{\Delta^2 k_\parallel dk_\parallel}{(k_\parallel^2 +\Delta^2)^3}\right\}\left\{ \int_0^\Lambda dk_\perp  k_\perp e^{-2\frac{k_\perp^2}{eB}}\right\}= \\
    \label{Fparallel}
    = 
    \begin{cases}
        \frac{1}{8\pi^2}\frac{eB}{\Delta^2}  & \textrm{if } eB \ll \Lambda^2 \\
        \frac{1}{4\pi^2}\frac{\Lambda^2}{\Delta^2}  & \textrm{if } \Lambda^2 \ll eB
    \end{cases}
\end{gather}
The parameters $Z_\varphi$ and $v_\perp^2$ now follow from $Z_\varphi^{-1} = F_\parallel$ and $v_\perp^2 = F_\perp/F_\parallel$. 

\bibliographystyle{JHEP}
\providecommand{\href}[2]{#2}\begingroup\raggedright\endgroup

\end{document}